\pdfoutput=1
\documentclass{sig-alternate-per}

\usepackage{graphicx}
\usepackage{epstopdf}
\usepackage{url}
\usepackage{mdwlist}
\usepackage{array}
\usepackage{multirow} 

\begin{document}

\title{DoKnowMe: Towards a Domain Knowledge-driven Methodology for Performance Evaluation}

\numberofauthors{3} 
\author{
\alignauthor Zheng Li\\
       \affaddr{Department of Electrical and Information Technology}\\
       \affaddr{Lund University}\\
       \affaddr{Lund, Sweden}\\
       \email{zheng.li@eit.lth.se}
\alignauthor Liam O'Brien\\
       \affaddr{ICT Innovation and Services}\\
       \affaddr{Geoscience Australia}\\
       \affaddr{Canberra, Australia}\\
       \email{liamob99@hotmail.com}
\alignauthor Maria Kihl\\ 
       \affaddr{Department of Electrical and Information Technology}\\
       \affaddr{Lund University}\\
       \affaddr{Lund, Sweden}\\
       \email{maria.kihl@eit.lth.se}
}

\maketitle
\begin{abstract}
Software engineering considers performance evaluation to be one of the key portions of software quality assurance. Unfortunately, there seems to be a lack of standard methodologies for performance evaluation even in the scope of experimental computer science. Inspired by the concept of ``instantiation" in object-oriented programming, we distinguish the generic performance evaluation logic from the distributed and ad-hoc relevant studies, and develop an abstract evaluation methodology (by analogy of ``class") we name Domain Knowledge-driven Methodology (DoKnowMe). By replacing five predefined domain-specific knowledge artefacts, DoKnowMe could be instantiated into specific methodologies (by analogy of ``object") to guide evaluators in performance evaluation of different software and even computing systems. We also propose a generic validation framework with four indicators (i.e.~usefulness, feasibility, effectiveness and repeatability), and use it to validate DoKnowMe in the Cloud services evaluation domain. Given the positive and promising validation result, we plan to integrate more common evaluation strategies to improve DoKnowMe and further focus on the performance evaluation of Cloud autoscaler systems. 
\end{abstract}





\section{Introduction}
\label{sec:introduction}

Performance evaluation has belonged to the experimental computer science field since the beginning of 1980s \cite{Denning_1981}. Unfortunately, compared to the well-established
theory and practice, computer science has put less focus on experimental
methodologies and scientific observations \cite{Feitelson_2006}. Although experimental computer science requires more standardization, researchers and practitioners in this field seem to have to borrow methodologies from natural sciences \cite{Feitelson_2006} and even from economics \cite{Grossklags_2007}, not to mention having a specific methodology for performance evaluation of software systems. In fact, the literature shows that the existing relevant studies usually employ ad-hoc approaches instead of strict means to implement performance evaluations (e.g., \cite{Li_Zhang_2013_slr}). In extreme cases, the evaluation methodology was treated as experimental setup and/or preparation of experimental environment; some evaluators only focused on metrics; while some others only highlighted benchmarks when specifying their evaluation approach. Thus, a natural question would be: Is there a dedicated methodology for software system performance evaluation? In general:

\begin{table}[h]\small
\centering
\begin{tabular}{p{7.6cm}}
\textbf{A methodology refers to ``an organised set of principles which
guide action in trying to `manage' (in the broad sense) real-world
problem situations." \protect\cite{Checkland_Scholes_1999}}  \\
\end{tabular}
\end{table}

When it comes to the performance evaluation problem, however, it seems impossible to obtain a one-size-fits-all methodology for evaluating performance of all kinds of software systems. Although there exist standard principles for guiding particular activities involved in performance evaluation (e.g., using randomization to eliminate measurement bias), different system domains might come with domain-specific evaluation concerns and experiments, because the features and characteristics of different software systems could vary significantly (e.g., performance evaluation of component-based software systems \cite{Koziolek_2010} vs.~performance evaluation of database-oriented software systems \cite{Osman_Knottenbelt_2012}). 

\begin{figure}[!t]
\centering
\includegraphics[width=8.5cm]{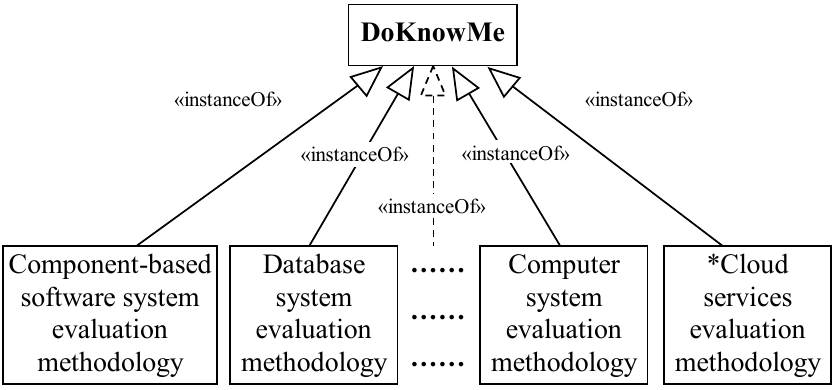}
\caption{The ``instantiation" relationship between DoKnowMe and its instance methodologies in different domains. (*Validations have been done in the Cloud services evaluation domain (cf.~Section \protect\ref{subsec:validatingCEEM}).)}
\label{fig_Instantiation}
\end{figure}

Inspired by the concept of \textit{Instantiation} in object-oriented programming, we developed a generic performance evaluation methodology that can be instantiated into various concrete methodologies for evaluating different software systems and even other computing systems, by integrating their corresponding domain-specific knowledge. We name this root methodology ``class" as Domain Knowledge-driven Methodology for Performance Evaluation (DoKnowMe), and all the potential concrete methodologies are its instance ``objects", as shown in Figure \ref{fig_Instantiation}. 

In addition to relying on our own experience, the development of DoKnowMe is largely based on two relevant disciplines. Firstly, it borrows common lessons from the existing performance evaluation work in computer science (e.g., \cite{Fortier_Michel_2003,Jain_1991,Le_Boudec_2011,Koziolek_2010,Osman_Knottenbelt_2012}). Secondly, it refers to the guidelines from Design of Experiments (DOE) \cite{Montgomery_2009}. Although DOE is normally applied to agriculture, chemical, and process industries, considering the natural relationship between experimentation and evaluation in this case, the various DOE techniques of experimental design and statistical analysis can also benefit performance evaluation of software/computing systems. In particular, as suggested in \cite{Feitelson_2015}, we define a set of knowledge artefacts integrated in DoKnowMe to organize and facilitate utilizing the domain-specific knowledge.

This paper introduces DoKnowMe that has been partially validated in the Cloud services evaluation domain. Note that although the concept of performance evaluation could include both experimental measurement and model-based prediction, DoKnowMe is defined to be applicable to the experimental measurement scenario only, while not to the model-based prediction scenario. We consider experimental performance measurement to be a fundamental evaluation scheme, because the model-based prediction still needs measurements to determine the performance annotations involved in the model \cite{Koziolek_2010}. 

The contributions of this work are mainly threefold.
\begin{itemize*}
	\item	We have developed a generic performance evaluation methodology, namely DoKnowMe, in the field of experimental computer science. To the best of our knowledge, DoKnowMe is the first unified methodology that can help guide performance evaluation for different software/computing systems through instantiation.  
	\item	Common experimental artefacts would be necessary and beneficial for sharing evaluation experiences. However, the existing studies mainly focused on repeating or reproducing previous experiments \cite{Feitelson_2006,Feitelson_2015}. We have defined five artefacts to cater for replaceable domain-specific knowledge. These knowledge artefacts are essentially used to instantiate DoKnowMe to facilitate evaluation implementations in a specific system domain. 
	\item	We have proposed a methodology validation framework covering four indicators, i.e. \textit{Usefulness}, \textit{Feasibility}, \textit{Effectiveness} and \textit{Repeatability}. In addition to using the four indicators to validate DoKnowMe in the Cloud service system domain, we argue that this validation framework would be able to guide methodology validation in a generic sense.
\end{itemize*}

The remainder of this paper is organized as follows. Section \ref{sec:artefacts} describes the five artefacts for sharing replaceable domain-specific knowledge of performance evaluation. Section \ref{sec:doKnowMe}
specifies the step-by-step procedure when applying DoKnowMe to performance evaluation implementations. A generic methodology validation framework is introduced in Section \ref{sec:validation}, and we also report our preliminary validations and the threats to validity of DoKnowMe at this current stage. Conclusions and some future work are discussed in Section \ref{sec:conclusion}.

\section{Domain Knowledge Artefacts}
\label{sec:artefacts}
As the name suggests, the domain knowledge plays a crucial role in our methodology. Domain-specific evaluation knowledge can be learnt from various publications and domain experts. However, the unstructured and distributed knowledge would be difficult for efficient reuse. Therefore, we define a set of knowledge artefacts to facilitate reusing the existing evaluation experiences, and they also need to be implemented on a case-by-case basis for different system domains. 

The knowledge artefacts integrated in DoKnowMe can be distinguished between pre-established and post-established. The pre-established artefacts include a \textit{Taxonomy}, a \textit{Metrics Catalogue}, an \textit{Experimental Factor Framework}, and a \textit{Conceptual Model} for building \textit{Experimental Blueprints}.

\begin{itemize*}

  \item	\textit{Taxonomy:} Considering that a well-founded taxonomy is significantly beneficial to corresponding research in any field of study \cite{Price_Baecker_1993}, the primary knowledge artefact is a taxonomy that collects, standardizes and organizes the relevant concepts and atomic elements of performance evaluation in a particular system domain. In particular, the taxonomy elements should at least include both the performance features to be evaluated, and the atomic scenes for setting up evaluation experiments. As such,
evaluators can conveniently investigate evaluation studies through a divide-and-conquer approach: the existing evaluation work can be analyzed through decomposition into elements for experience summarization, while new evaluation scenarios can be portrayed through composing relevant taxonomy elements. 
  \item	\textit{Metrics Catalogue (e.g.,~\cite{Li_OBrien_2012_metrics}):} According to the lessons from computer system evaluation, a performance evaluation study must choose a set of performance metrics \cite{Jain_1991}, and each metric provides a different lens into the performance \cite{Fortier_Michel_2003}. To facilitate choosing metrics, here we define a dictionary-type of knowledge artefact that enables using performance features as retrieval keys to look up suitable metrics and their corresponding benchmarks. Such a knowledge artefact, namely \textit{Metrics Catalogue}, can be established along a regression manner, by using the aforementioned divide-and-conquer approach. In other words, benefiting from the clarified performance features and experimental setup scenes in the taxonomy, we can realize a metric lookup capability by isolating and collecting the de facto metrics from the existing evaluation work in a particular system domain.
  \item	\textit{Experimental Factor Framework (e.g.,~\cite{Li_OBrien_2012_factor}):} In any system performance evaluation, a proper set of experiments must be designed, while the relevant factors that have impacts on performance play a prerequisite role in designing evaluation experiments \cite{Jain_1991,Montgomery_2009}. Similar to \textit{Metrics Catalogue}, an experimental factor framework is also a dictionary-like knowledge artefact built from the existing domain-specific evaluation studies. This artefact helps evaluators identify suitable experimental factors while excluding others in a concrete space instead of on the fly, which essentially indicates a systematic rather than an ad-hoc decision making process. Note that this factor framework is supposed to supplement, but not replace, the expert judgment for experimental factor identification, and it would be particularly helpful for performance evaluation when there is a lack of experts.
  \item	\textit{Conceptual Model (e.g.,~\cite{Li_OBrien_2014}):} Considering ``a picture is worth a thousand words", a domain-specific evaluation conceptual model can be viewed as a graphical extension of its corresponding taxonomy. It further rationalizes and emphasizes the detailed relationships among evaluation elements and classifiers, so as to portray and even characterize actual evaluation experiments. For complex evaluation projects involving collaboration between multiple evaluators, characterizing experiments is particularly helpful to facilitate information exchange and avoid experimental duplications. In practice, relevant elements and classifiers can be employed to conveniently compose natural languages-style descriptions together with UML-style \textit{Experimental Blueprints} for recording and sharing evaluation experimental design.
\end{itemize*}

The post-established artefact is a library of DoKnowMe-based \textit{Evaluation Templates}.
\begin{itemize*}
\item	\textit{Evaluation Templates:} When conducting a performance evaluation by different people at different times and locations, a common requirement would be reproducible evaluation implementations and comparable experimental results. DoKnowMe-driven evaluation maintains a live document that cannot only help deliver a structured report to the recipient of evaluation result, but also help generate evaluation templates to enhance the experimental repeatability. An evaluation template records the implementation details of evaluating a particular performance feature, including pre-experimental information, experimental instructions, and automated experimental actions. As such, evaluators can directly reuse suitable DoKnowMe-based templates to facilitate reproducing evaluation implementations and making experimental results more comparable. A template library can eventually be established by gradually accumulating DoKnowMe-based evaluation templates. 
\end{itemize*}

\section{Step-by-step Procedure of Using DoKnowMe}
\label{sec:doKnowMe}
\begin{figure*}[!t]
\centering
\includegraphics[width=17.1cm]{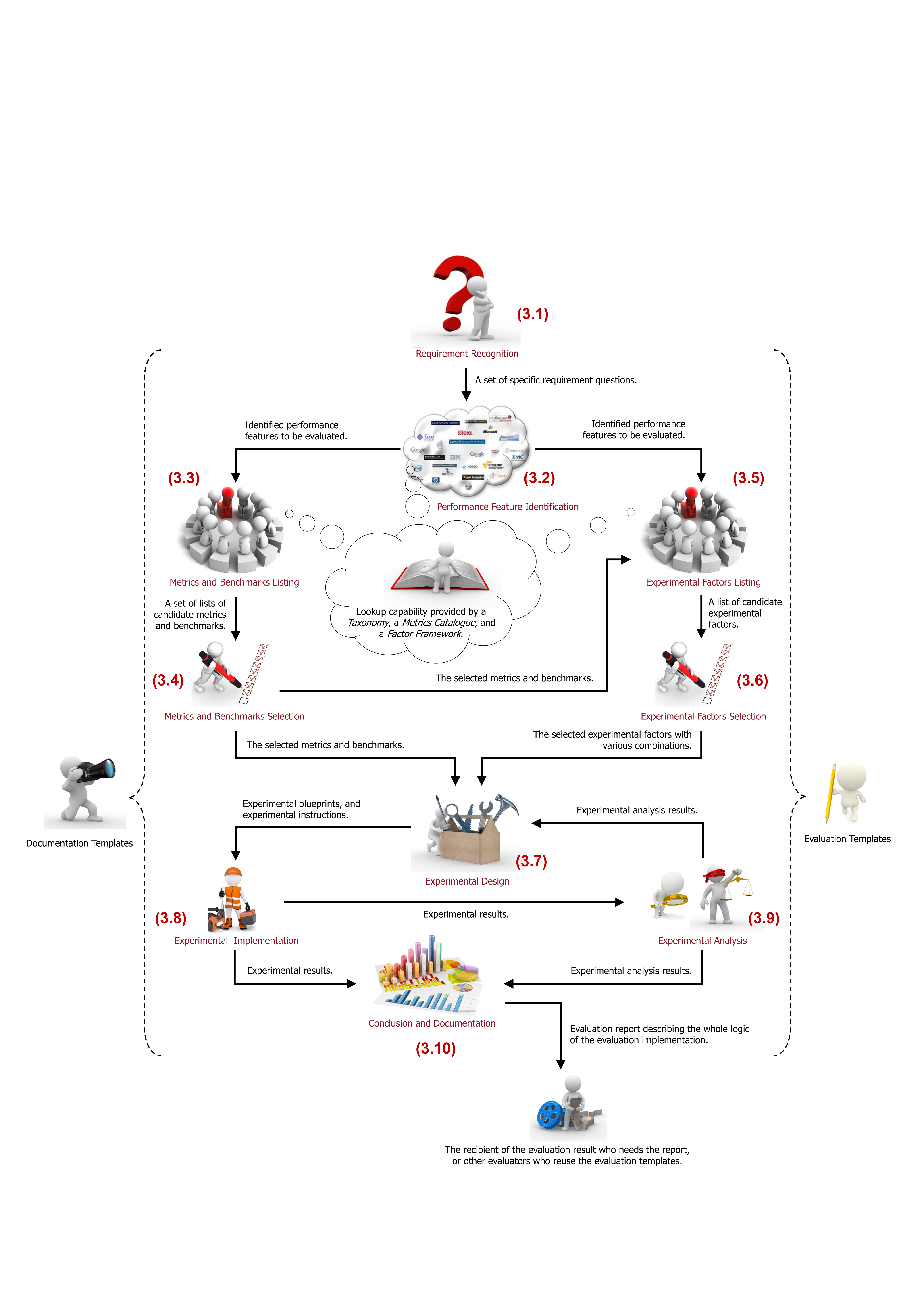}
\caption{The step-by-step procedure of using DoKnowMe.}
\label{fig_DoKnowMe}
\end{figure*}

The procedure of performance evaluation driven by DoKnowMe can be imagined as a sequential-process main thread, plus a possible spiral-process part representing recursive experimental activities. The recursive experimental activities will happen when an evaluation implementation is composed of a set of experiments, and the experimental design in a later iteration has to be determined by the experimental results and analyses from a prior iteration. Such a procedure can be illustrated as shown in Figure \ref{fig_DoKnowMe}.  

Inspired by the system thinking from the perspective of electrical dynamics, we further consider each evaluation step to be an input-output component of the whole methodology. A methodology component here essentially comprises a set of evaluation activities, and DoKnowMe particularly integrates strategies to facilitate conducting these activities. The strategies indicate both the generic evaluation lessons and the utilization of our knowledge artefacts. In this way, the evaluation steps can be described by using their inputs, activities with corresponding evaluation strategies, and outputs, 
as individually specified in the following subsections.

Note that this paper is not supposed to stick to any specific system domain, and thus reporting a detailed evaluation work is out of the scope of this paper. Therefore, we only adopt some conceptual examples to demonstrate particular evaluation steps involved in DoKnowMe.

\subsection{Requirement Recognition}
\label{subsec:requirement}
The recognition of an evaluation requirement is not only to understand a problem related to the system performance evaluation, but also to achieve a clear statement of the evaluation purpose, which is an obvious while non-trivial task \cite{Montgomery_2009}. A clearly specified evaluation requirement can facilitate driving the remaining steps properly in the evaluation implementation. To help recognize a requirement, it has been suggested to prepare a set of specific questions to be addressed by potential evaluation experiments \cite{Montgomery_2009}. Moreover, it is normally helpful to replace one comprehensive question with a list of separate and more easily answerable questions, so that evaluators can conveniently define specific evaluation objectives, and then employ the strategy of sequential experiments to satisfy the overall evaluation requirement. 

The elements of the step \textit{Requirement Recognition} are specified as below.

\noindent
\textit{\textbf{$<$Input:$>$}}

The input for recognizing an evaluation requirement is the natural-language description of a system evaluation problem. If an evaluation implementation is not in the charge of the recipients of evaluation results, the evaluation problem description will need to be generated by the discussions between the evaluators and the result recipients. 

\noindent
\textit{\textbf{$<$Activity:$>$}}

According to the natural-language description of a system performance evaluation problem, evaluators clarify and specify the objectives of the forthcoming evaluation implementation. Note that the clearly defined performance goals for a system would also be useful and necessary for clarifying its evaluation objectives.

\noindent
\textit{\textbf{$<$Strategy:$>$}}

The main strategy here is to use a standard terminology to define evaluation objectives into requirement questions. This can be realized by creating a mapping between the elements in a well-established \textit{Taxonomy} (e.g.,~a taxonomy for performance evaluation of Cloud services \cite{Li_OBrien_2014}) and the keywords in the evaluation problem description. If necessary, the natural-language problem description can be rephrased by using the relevant taxonomy elements in advance.   

Moreover, evaluators need to revise the requirement questions to make them as specific as possible, for example, breaking one comprehensive question into a list of separate and more easily answerable questions.

\noindent
\textit{\textbf{$<$Output:$>$}}

The recognition output is a set of specific requirement questions to be addressed by potential evaluation experiments.

\noindent
\textit{\textbf{Example:}}

Suppose users are concerned with a modern software system that is ``expected to deliver reliable performance under highly variable load intensities" \cite{Kistowski_Herbst_2014}. Although the performance keywords in this natural-language description are ``reliable" and ``variable", the real performance concern should be how well the software system scales, according to the specifications of reliability, variability, and scalability \cite{Li_OBrien_2014}. Correspondingly, we can define three specific requirement questions for driving this software system performance evaluation, such as:

\begin{itemize*}
  \item	How scalable is the software system when dealing with different amounts of workloads?
  \item	How fast does the software system scale with an increasing workload?
  \item	How fast does the software system scale with a decreasing workload?
\end{itemize*}

\subsection{Performance Feature Identification}
Given the clarified evaluation requirement of a system, evaluators need to further identify relevant performance features to be evaluated. Since different end users could be interested in numerous quality aspects of various functionalities of a particular software/computing system, it would be difficult for evaluators to directly locate proper performance features. Therefore, it will be helpful and valuable to have a relatively complete set of domain-specific feature candidates in advance. Then, in most cases, suitable performance features can be conveniently selected from the candidate feature set. 

By standardizing the terms/concepts and their relationships (e.g., the whole-part relationship) within a system domain, the knowledge artefact \textit{Taxonomy} essentially provides such a set of candidate performance features with respect to a family of systems, which correspondingly facilitates \textit{Performance Feature Identification}, as specified below. Note that the widely-used term ``feature" itself also needs to be clarified in the domain-specific \textit{Taxonomy}. For example, a particular Cloud service's performance feature can be defined as a combination of the service's physical property and its capacity \cite{Li_OBrien_2014}.

\noindent
\textit{\textbf{$<$Input:$>$}}

Following \textit{Requirement Recognition}, the input here is a set of specific evaluation requirement questions. 

\noindent
\textit{\textbf{$<$Activity:$>$}}

Evaluators rescan the predefined requirement questions, and identify relevant performance features from each of them. 

\noindent
\textit{\textbf{$<$Strategy:$>$}}

Evaluators still utilize the pre-established \textit{Taxonomy} to explain the system quality concerns from each of the requirement questions. The performance features to be evaluated can then be determined by matching the quality concern explanations with the performance feature descriptions.

If possible, evaluators can further consult domain experts to help verify and identify performance features. 

\noindent
\textit{\textbf{$<$Output:$>$}}

This step outputs the identified performance features to be evaluated. 

\noindent
\textit{\textbf{Example:}}

Recall that the first requirement question focuses on the capacity of eventually catering the changed workload, while the other two questions emphasizes the speed of response to the changing for workload. By exploring the domain knowledge, It is clear that such explanations match the definitions of scalability and elasticity respectively \cite{Islam_Lee_2012}. Therefore, the three requirement questions correspond to two features of the software system, i.e., scalability and elasticity. 

\subsection{Metrics and Benchmarks Listing}
\label{chap4:subsec:metric_listing}
The choice of the right metrics depends on the identified performance features to be evaluated \cite{Jain_1991}, while given that the performance features are yet insufficient for choosing right metrics. On the one hand, a performance feature can be measured by different metrics with different benchmarks \cite{Li_OBrien_2012_metrics}; on the other hand, the selection of particular metrics and benchmarks might have other constraints or tradeoffs (see the step \textit{Metrics and Benchmarks Selection} below). Therefore, to facilitate the metric/benchmark selection, it is also helpful for evaluators to refer to the existing available metrics and benchmarks for the corresponding evaluation implementation. 

The purpose of the step \textit{Metrics and Benchmarks Listing} is thus to list as many as possible candidate metrics and benchmarks for evaluating the identified performance features. The elements of this step are specified as below.

\noindent
\textit{\textbf{$<$Input:$>$}}

The input is the identified performance features to be evaluated. 

\noindent
\textit{\textbf{$<$Activity:$>$}}

For each of the identified performance features, evaluators enumerate as many relevant metrics and benchmarks as they can come up with.

\noindent
\textit{\textbf{$<$Strategy:$>$}}

Evaluators can use the identified Cloud service features as retrieval keys to quickly search metrics and benchmarks from the knowledge artefact \textit{Metrics Catalogue} (e.g.,~a metrics catalogue for
performance evaluation of Cloud services \cite{Li_OBrien_2012_metrics})\footnote{An online version of this metrics catalogue can be found at: \url{http://cloudservicesevaluation.appspot.com/}}. The \textit{Metrics Catalogue} essentially provides a lookup capability for finding candidate metrics and benchmarks from the existing evaluation experiences in a particular system domain. 

Note that group meetings and expert opinions are not supposed to be completely replaced with this \textit{Metrics Catalogue}. New metrics and benchmarks can still be introduced by domain experts and other evaluators.

\noindent
\textit{\textbf{$<$Output:$>$}}

The output of this step is a set of lists of candidate metrics and benchmarks, and each list is related to a single performance feature. 

\noindent
\textit{\textbf{Example:}}

In general, scalability has to be reflected by the change of value of some primary performance features \cite{Li_OBrien_2014}. To simplify the demonstration, suppose we measure the primary feature ``communication data throughput" to reflect the previously identified feature scalability. Benefiting from the pre-built metrics catalogue \cite{Li_OBrien_2012_metrics} (note that this catalogue is for evaluating Cloud service systems only), we can quickly list two candidate metrics and each of them corresponds to four candidate benchmarks for evaluating the performance feature ``communication data throughput", as shown in Table \ref{tbl>communicationMetrics}. 

\begin{table}[!t]\small 
\renewcommand{\arraystretch}{1.3}
\centering
\caption{\label{tbl>communicationMetrics}Candidate Metrics and Benchmarks for Evaluating Communication Data Throughput (originally appeared in {\protect\cite{Li_OBrien_2012_metrics}})}

\begin{tabular}{|p{2cm} |>{\raggedright}p{2cm}| >{\raggedright\arraybackslash}p{3.1cm}|}
\hline

\textbf{Feature} & \textbf{Metric} & \textbf{Benchmark}\\
\hline
\multirow{9}{2cm}{Communication Data Throughput} & \multirow{5}{2cm}{TCP/UDP/IP Transfer Speed} & iPerf\\
\cline{3-3}
& & Private tools TCPTest/UDPTest \\
\cline{3-3}
& & SPECweb 2005 \\
\cline{3-3}
& & Upload/Download/Send large size data\\
\cline{2-3}
 & \multirow{4}{2cm}{MPI Transfer Speed} & HPCC: b\_eff \\
\cline{3-3}
& & Intel MPI Bench  \\
\cline{3-3}
& & mpptest \\
\cline{3-3}
& & OMB-3.1 with MPI  \\
\hline

\end{tabular}
\end{table}

\subsection{Metrics and Benchmarks Selection}
\label{subsec:metric_selection}
A metric is a measurable quantity that precisely captures some characteristics of a performance feature. According to the rich research in the evaluation of computer systems, the selection of metrics plays an essential role in evaluation implementations \cite{Obaidat_Boudriga_2010}. Furthermore, a suitable metric would play a \textit{Response Variable} role \cite{Montgomery_2009} in applying DOE to performance evaluation. Although traditional evaluation lessons treat metrics selection as one of the prerequisites of benchmark selection \cite{Jain_1991}, the availability of benchmarks could in turn constrain the employment of metrics. For example, if adopting the benchmark iPerf, only one metric (i.e.~\textit{TCP/UDP/IP Transfer Speed}) can be selected from the list to measure the feature communication data throughput (cf.~Table \ref{tbl>communicationMetrics}). Therefore, DoKnowMe puts the selection of metrics and benchmarks together within one step.

The elements of this step are specified as below. Since the selection of metrics and benchmarks is fairly straightforward based on pre-listed candidates, we do not use any example to demonstrate this step.

\noindent
\textit{\textbf{$<$Input:$>$}}

This step takes the lists of candidate metrics and benchmarks as input. 

\noindent
\textit{\textbf{$<$Activity:$>$}}

Evaluators select the most appropriate metrics and benchmarks from the pre-listed candidates. In addition, evaluators can also develop benchmark tools by themselves if there is a lack of suitable resources.

\noindent
\textit{\textbf{$<$Strategy:$>$}}

Based on the candidate lists, the decision on metrics and benchmarks selection can be made by checking the available resources in hand, estimating the overhead of potential experiments, and judging the evaluators' capabilities of operating different benchmarks.

When it is necessary to measure various performance features from the holistic perspective, evaluators can further employ the \textit{Boosting Metrics} technique \cite{Li_OBrien_2013_boosting} to combine various indicators into a single index. In brief, a boosting metric is a secondary measurement
criterion by manipulating a set of primary metrics that
directly measure individual performance features.

\noindent
\textit{\textbf{$<$Output:$>$}}

This step outputs the selected metrics and benchmarks.

\subsection{Experimental Factors Listing}
\label{subsec:factorlisting}
To evaluate a performance feature, knowing all factors (also called parameters or variables) that affect the performance feature has been considered to be a tedious but crucial task \cite{Le_Boudec_2011}. Although listing a complete scope of experimental factors may not be easily achieved, at all times evaluators should keep the factor list as comprehensive as possible, for further analysis and decision making about the factor selection and data collection \cite{Jain_1991}. 

Thus, the purpose of \textit{Experimental Factors Listing} is to list all the candidate experimental factors related to the performance features to be evaluated. The elements of this step are specified as below.

\noindent
\textit{\textbf{$<$Input:$>$}}

The input here has two parts: the first part is the identified performance features to be evaluated, and second part is the selected metrics and benchmarks. 

\noindent
\textit{\textbf{$<$Activity:$>$}}

According to the identified performance features and the selected metrics and benchmarks, evaluators list potential candidate experimental factors.

\noindent
\textit{\textbf{$<$Strategy:$>$}}

Evaluators can screen and lookup potential factors in the knowledge artefact \textit{Experimental Factor Framework} (e.g.,~a factor framework for performance evaluation of Cloud services \cite{Li_OBrien_2012_factor}). In particular, the identified performance features are used to explore system-contained and/or environmental Resource factors, the selected benchmarks are used to search Workload factors, and the selected metrics are directly used as system Quality factors. Note that following DoKnowMe evaluators only need to list input-process factors (i.e.~resource and workload factors), because the output-process factors (i.e.~system quality factors) are essentially the metrics that have been selected in the previous step.

If possible, evaluators can further hold group discussions and consult domain experts for listing potential factors and updating the \textit{Experimental Factor Framework}. Although an ideal factor framework is supposed to capture the state-of-the-practice of experimental factors in a particular system domain, the ``state of the practice" does not imply that the framework has exhaustively provided all the potential factors. Instead, this knowledge artefact offers a concrete and rational basis for further discussion and factor listing by expert judgments. 

\noindent
\textit{\textbf{$<$Output:$>$}}

The output includes three types of candidate experimental factors, with respect to the identified performance feature-related resource, workload and system quality respectively. 

\noindent
\textit{\textbf{Example:}}

\begin{figure}[!t]
\centering
\includegraphics[width=7cm]{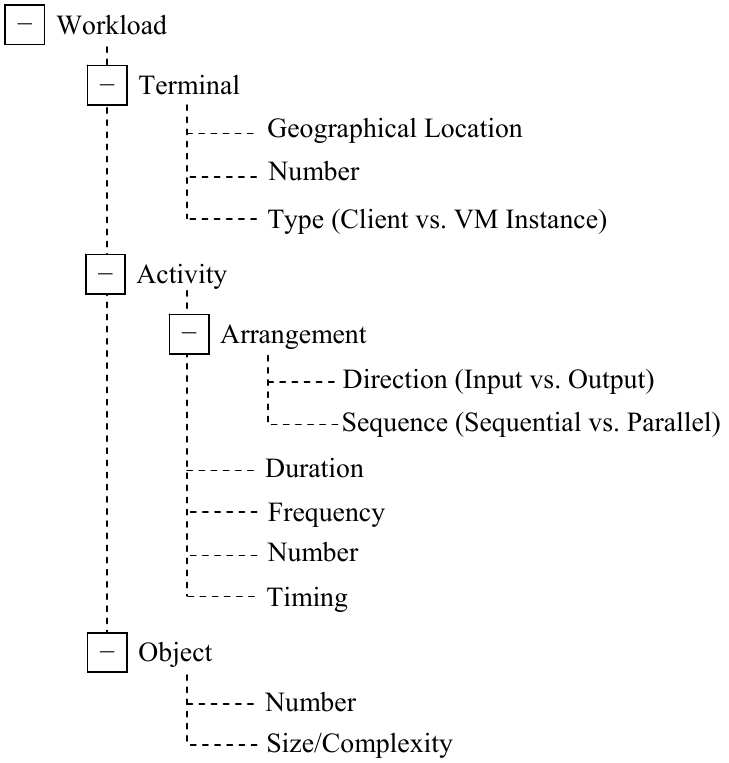}
\caption{Workload-related factors (originally appeared in \protect\cite{Li_OBrien_2012_factor}).}
\label{fig_WorkloadTree}
\end{figure}

Recall that the evaluation requirement example is mainly concerned with highly variable workloads, we particularly demonstrate potential workload-related factors, as illustrated in Figure \ref{fig_WorkloadTree}. Our previous work shows that workload used in performance evaluation could be described through one of three different concerns or a combination of them, namely Terminal, Activity, and Object. More details have been specified in \cite{Li_OBrien_2012_factor}.

\subsection{Experimental Factors Selection}
It is clear that the determination of factors and their levels/ranges is the prerequisite of a factor-based experimental design \cite{Montgomery_2009}. Furthermore, it is better to start with limited design factors distinguished from nuisance ones and those that are not of interest, and the factors that are expected to have high impacts should be preferably selected \cite{Jain_1991}. Note that there is no conflict between selecting limited factors in this step and keeping a comprehensive factor list in the previous step. On the one hand, an evaluation requirement usually comprises a set of experiments, and different experiments might have to select different factors from the same factor list. On the other hand, intentionally excluding unused factors does not mean that evaluators have not considered them, and some of the excluded factors can also be saved for reinforcement experiments in the future.

The elements of \textit{Experimental Factors Selection} are specified as below.

\noindent
\textit{\textbf{$<$Input:$>$}}

The input of this step is a list of candidate experimental factors. 

\noindent
\textit{\textbf{$<$Activity:$>$}}

Given the pre-listed candidate experimental factors, evaluators select a limited set that are of most interest, and also determine the values or ranges of the selected factors.

\noindent
\textit{\textbf{$<$Strategy:$>$}}

Suitable experimental factors can be selected by roughly considering pilot experiments without detailed experimental design. Following DoKnowMe, evaluators can use the experimental setup scenes clarified in the \textit{Taxonomy} to try making up different experimental scenarios by combining candidate factors. Through this way, the factor selection eventually becomes determining experimental scenarios that have been constrained by the pre-specified evaluation requirement questions, as demonstrated in Section \ref{subsec:requirement}. Driven by separate experimental scenarios, different combinations of selected factors would be either independently involved in parallel experimental processes, or incrementally involved in a consecutive and/or recursive experimental process. 

\noindent
\textit{\textbf{$<$Output:$>$}}

The output is one or more groups of selected experimental factors, corresponding to various factor combinations for designing different experiments.

\subsection{Experimental Design}
\label{chap4:subsec:design}
Once experimental factors are selected, evaluation experiments can subsequently be prepared and designed. Normally, a small scale of pilot experiments would benefit the relevant experimental design, for example, by helping evaluators get familiar with the experimental environment, optimize the experimental sequence, etc.

The elements of \textit{Experimental Design} are specified as below.

\noindent
\textit{\textbf{$<$Input:$>$}}

The input here includes the selected metrics, benchmarks and experimental factors, and possible results and analyses from trial experiments. 

\noindent
\textit{\textbf{$<$Activity:$>$}}

DoKnowMe recognizes four consecutive activities during experimental design:
\begin{itemize*}
  \item	Evaluators design simple experiments based on pilot trials which includes developing codes/scripts for automatically preparing environments and driving benchmarks.
  \item	Evaluators use DOE techniques to design more complex experiments.
  \item	If necessary, evaluators modify experiments based on previous iteration experimental results and analyses. 
  \item	Evaluators record and even characterize experiments for delivering the experimental design.
\end{itemize*}

\noindent
\textit{\textbf{$<$Strategy:$>$}}

Designing simple experiments is based on pilot trials that act as a continuation of experimental factor selection. As previously mentioned, pilot trials are essentially the implementations of different experimental scenarios, while the experimental scenarios can be generated by using suitable experimental setup scenes listed in the \textit{Taxonomy}.

When it comes to designing complex experiments, evaluators can select suitable DOE techniques integrated in DoKnowMe. For example, using operating characteristic (OC) curves to decide the sample size -- number of replicates, or using full factorial design to facilitate identifying the most influential factor. In particular, three basic principles of DOE, namely Randomization, Replication, and Blocking \cite{Montgomery_2009}, should be generally taken into account when designing experiments. 

The designed experiments can be further characterized and recorded by building UML-style \textit{Experimental Blueprints}. As a supplement to literal experimental instructions, evaluators can use experimental blueprints to facilitate discussions among different people, or facilitate ``apple-to-apple" comparison between different evaluation experiments. 

\noindent
\textit{\textbf{$<$Output:$>$}}

The output includes experimental instructions, experimental blueprints, and codes/scripts for preparing experiment environment and driving benchmarks. 

\noindent
\textit{\textbf{Example:}}

\begin{figure}[!t]
\centering
\includegraphics[width=7cm]{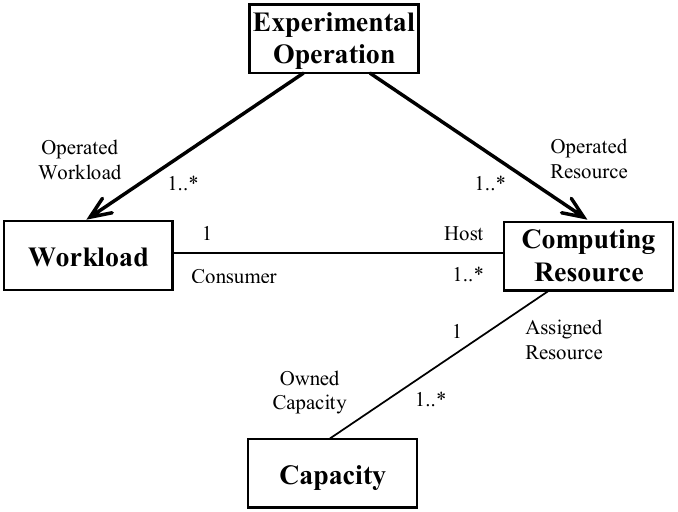}
\caption{A root blueprint for evaluating a computing system (originally appeared in \protect\cite{Li_OBrien_2014}).}
\label{fig_blueprint}
\end{figure}

Due to the space limit, here we only demonstrate a root blueprint at the top abstract level (cf.~Figure \ref{fig_blueprint}) for evaluating a computing system. As explained in \cite{Li_OBrien_2014}, this blueprint reflects the most generic reality of a system performance evaluation, i.e., ``evaluating \textit{[capacity]} of particular \textit{[computing resources]} with particular \textit{[workload]} driven by a set of \textit{[experimental operations]}".

\subsection{Experimental Implementation}
\label{subsec:implementation}
Implementing an experiment involves carrying out a series of experimental actions ranging from preparing experiment environment to running benchmarks. Since any error in the experimental procedure may spoil the validity of the experimental results, the implementation process should be monitored carefully to ensure that every detail of the experiments follows the design \cite{Montgomery_2009}. 

To help guide \textit{Experimental Implementation}, DoKnowMe emphasizes specifying and automating experimental actions (e.g.,~repeat running benchmark for 24 hours) based on the design, and the elements of this step are specified as below.

\noindent
\textit{\textbf{$<$Input:$>$}}

The input of this step accepts the experimental design materials. 

\noindent
\textit{\textbf{$<$Activity:$>$}}

Evaluators record experiment environment (e.g., time and location), carry out the designed experiments and obtain experimental results. Based on their practical activities, evaluators further improve the automation of experimental actions.

\noindent
\textit{\textbf{$<$Strategy:$>$}}

DoKnowMe requires experimental implementations to rigorously follow their corresponding design. To reach the rigorousness, evaluators can make the experimental actions automated as much as possible to increase repeatability and reduce human mistakes. Note that DoKnowMe regards pilot experimental runs as the evaluation activities in \textit{Experimental Design} instead of any activity in this step.

\noindent
\textit{\textbf{$<$Output:$>$}}

This step not only outputs experimental results, but also delivers automation codes/scripts for experimental implementation.

\subsection{Experimental Analysis}
\label{chap4:subsec:analysis}
As demonstrated so far, experimental results can sometimes answer their corresponding evaluation requirement questions already. However, in most cases, more convincing conclusions would have to be drawn through further experimental analysis. In general, an experimental analysis could heavily rely on statistical methods \cite{Montgomery_2009}. Although such methods do not directly prove any factor's effect, the statistical analysis adds objectivity to drawing evaluation conclusions and potential decision-making process.  

The elements of \textit{Experimental Analysis} are specified as below.

\noindent
\textit{\textbf{$<$Input:$>$}}

The input is the experimental results. 

\noindent
\textit{\textbf{$<$Activity:$>$}}

Evaluators use statistical techniques to analyze experimental results, and visualize both experimental results and analysis results if applicable.

\noindent
\textit{\textbf{$<$Strategy:$>$}}

Being compatible with the traditional evaluation lessons, DoKnowMe also emphasizes that visualizing experimental results by using various graphical tools would significantly facilitate data analysis and interpretation. In our existing practices, suitable charts are directly used as metrics for measuring performance, as defined in the knowledge artefact \textit{Metrics Catalogue} \cite{Li_OBrien_2012_metrics}. For example, representations in column, line, and scatter can be used as scalability and variability evaluation metrics; while the radar plot is used as one of the typical preliminary boosting metrics \cite{Li_OBrien_2013_boosting}. 

When it comes to some complex cases, similarly, the evaluators can employ suitable DOE techniques integrated in DoKnowMe to analyze experimental results. For example, using Analysis of Variance (ANOVA) to reveal unclear characteristics of variability, or using Pareto plot to illustrate the performance effects of different factors and factor combinations \cite{Li_OBrien_2012_factor}. 

\noindent
\textit{\textbf{$<$Output:$>$}}

Similarly, the output here includes not only experimental analysis results but also potential codes/scripts for automating experimental analysis.

\subsection{Conclusion and Documentation}
Drawing appropriate conclusions is significant after analyzing the experimental results \cite{Montgomery_2009}. In addition, it is worth paying more attention to reporting the whole performance evaluation work \cite{Heiser_2014}. In fact, not only conclusions but also complete evaluation reports would be vital for other people to learn from or replicate/confirm previous evaluation practices.

DoKnowMe uses a structured manner to implement \textit{Conclusion and Documentation}, and the elements of this step are specified as below.

\noindent
\textit{\textbf{$<$Input:$>$}}

The input includes both the experimental results and the experimental analysis results. 

\noindent
\textit{\textbf{$<$Activity:$>$}}

Evaluators build mappings between the experimental (analysis) results and the requirement questions. Furthermore, in most cases, evaluators summarize conclusions based on the experimental results and analysis results.

By this step, evaluators finalize documenting the evaluation study. Note that documentation is a default activity at each step of DoKnowMe.

\noindent
\textit{\textbf{$<$Strategy:$>$}}

Evaluators can directly use tables and visual representations of experimental (analysis) results to respond to the pre-specified requirement questions. 
The answers to all the requirement questions can further be summarized into natural-language findings to better reflect the conclusions. 

As for the documentation, evaluators can follow the steps of DoKnowMe to generate structured evaluation reports and evaluation templates. In particular, the evaluation report mainly focuses on the whole logic of evaluation procedure in natural language, while the evaluation templates mostly record the detailed environmental information, experimental instructions, and automated experimental actions to facilitate evaluation replications. Generally, an evaluation implementation would deliver a set of different templates for evaluating different performance features. 
 
Although evaluators can employ any well-proposed technique of reporting experiments or case studies \cite{Runeson_2009} to help document performance evaluation studies, the advantage of following DoKnowMe is that, by recording the evaluation activities of every single step, a \textit{live and structured log} is essentially maintained and self-documented. This self-documented log can then act as a base for generating evaluation report and DoKnowMe-based evaluation templates.

\noindent
\textit{\textbf{$<$Output:$>$}}

The final step of DoKnowMe delivers evaluation conclusions, a complete evaluation report, and a set of DoKnowMe-based evaluation templates. 

\section{Validating DoKnowMe}
\label{sec:validation}
\subsection{A Generic Validation Framework}

To help validate a performance evaluation methodology in a systematic way, we propose a generic validation framework with four indicators:

\begin{itemize*}
  \item	Usefulness
  \item	Feasibility
  \item	Effectiveness 
  \item	Repeatability
\end{itemize*}

Given the definition of methodology (cf.~Section \ref{sec:introduction}), the indicator \textit{Usefulness} implies that the methodology to be validated should be helpful for guiding human activities to deal with real-world problems. In this case, a methodology would be useful as long as it can drive performance evaluation implementations. 

However, a useful evaluation methodology is not necessarily feasible. Compared to the existing performance evaluation approaches, a new methodology is feasible only when it has more advantages in evaluating performance. In other words, an alternative methodology might not be worth being employed even if it works as well as the others. Therefore, we further use \textit{Feasibility} to indicate the validation scenario of replicating the existing studies, to check if the useful methodology can bring more value than the existing approaches to performance evaluation.

Compared to feasibility studies that are used to assess whether or not the ideas and findings are appropriate for further testing, effectiveness studies are used to evaluate success in real-world with non-ideal conditions \cite{Bowen_Kreuter_2009}. Therefore, we use \textit{Effectiveness} to indicate the validation scenario of applying the feasible methodology to new performance evaluation problems with real requirements. 

When it comes to \textit{Repeatability}, we further emphasize two views of repeatedly applying a methodology. The first view is to validate whether or not an alternative methodology is repeatable for reproducing an evaluation implementation, from the dynamic perspective of evaluation activities. The second view is to validate whether or not an alternative methodology can lead to comparable results of the corresponding activities, from the static perspective of evaluation outputs. Overall, given the same performance evaluation requirement and the same methodology, we are concerned whether or not different evaluators can independently work along (at least nearly) the same evaluation processes and generate (at least nearly) the same outputs at each evaluation step.

\subsection{Validating DoKnowMe by Validating its Instance Methodology}
\label{subsec:validatingCEEM}
Since DoKnowMe is a root methodology for performance evaluation of different types of software/computing systems, the validation of DoKnowMe has to be reflected by validating its instance methodologies. By instantiating DoKnowMe, we have developed and validated a Cloud Evaluation Experiment Methodology (CEEM), and correspondingly validated DoKnowMe in the Cloud services evaluation domain.

In detail, to validate the usefulness and feasibility of CEEM, we performed two case studies of replicating past Cloud services evaluation implementations (computation evaluation of Google AppEngine Python runtime and storage evaluation of different types of Amazon EC2 instances), and compared our practices with the original ones. The results show that our replication work followed a more systematic and complete procedure driven by CEEM, and the systematic evaluation procedure brought more value than those original
studies. For example, our practices delivered more convincing results and conclusions by rigorously conducting every single evaluation step, and the integrated DOE techniques helped reveal more information behind the
Cloud service specifications.

To verify the effectiveness and repeatability of CEEM, we performed four case
studies of implementing new Cloud services evaluations: (1) Selection between two similar types of Amazon EC2 instances; (2) An ``apple-to-apple" comparison of computation performance between Google Compute Engine (GCE) and Amazon EC2; (3) Communication, memory, storage and computation evaluations of the recently available GCE; and (4) A relatively complicated project of evaluating content delivery network applications over public Clouds by different evaluators. These four case studies show that CEEM can effectively drive evaluation of Cloud services to satisfy real-world requirements. Each case study has followed the CEEM-based documentation structure to report the evaluation implementation details. The integrated knowledge artefacts have been particularly validated as valuable and helpful for facilitating both relevant evaluation activities and expert judgments. Moreover, it is clear that CEEM is repeatable for evaluating and comparing different Cloud services, even some characteristics of the evaluated services are fairly different. When employing CEEM independently by different people, the fourth case study shows that different evaluators can obtain similar outputs at each evaluation step within the same requirement domain. More importantly, the CEEM-based evaluation templates can be conveniently reused to significantly facilitate new evaluation implementations.

\subsection{Threats to Validity}
As mentioned previously, a root methodology cannot directly be validated. To justify the general applicability of DoKnowMe to different types of systems, we need to either exhaustively validate its instance methodologies in all system domains, or we should realize the general validation through an induction approach, i.e.~by reasoning a set of individual domain-specific validation instances. Considering that we have only conducted validations in the Cloud services evaluation domain, DoKnowMe is still premature at this current stage, and its validation requires instantiating and verifying additional methodologies for other software system types.

To relieve the threat to our work's validity, we are currently working with several research groups on applying DoKnowMe to their performance evaluation studies. For example, DoKnowMe has been involved in an autoscaler evaluation project by our colleagues from Ume\r{a} University. Autoscalers are a special type of software that aims for efficient resource management through automatic scaling in Cloud computing. We plan to reuse the successful experience of evaluating Cloud services, and now we are developing the autoscaler-oriented knowledge artefacts. Another example is that DoKnowMe was just employed for evaluating machine learning tools by two researchers at Concordia University. Overall, it is clear that applying our methodology to various system domains will need numerous performance evaluation practices that might require extensive collaborations with different researchers and practitioners. Therefore, we try to broadcast DoKnowMe to speed up its applicability verification.

\section{Conclusions}
\label{sec:conclusion}
It has been identified that performance plays a key role in the success of software systems, and its evaluation is a crucial approach to assuring software performance. Thus, a suitable methodology would be necessary and helpful to guide evaluators in implementing performance evaluations. Considering numerous and various software features and characteristics in different domains, it could be impossible to come up with a universal evaluation methodology for measuring performance of different types of software systems. It is also irrational to build individual evaluation methodologies from scratch for different systems, because there exist common principles for some generic evaluation steps and activities.

By distinguishing the domain-independent logic of performance evaluation from the distributed and ad-hoc relevant studies, we developed a domain knowledge-driven evaluation methodology, namely DoKnowMe, to act as a unified methodology in experimental computer science. Following the idea of ``instantiation" in object-oriented thinking, it would be possible to instantiate DoKnowMe by integrating domain-specific knowledge artefacts to facilitate evaluating different software and even computing systems. 

After preliminarily validating DoKnowMe in the Cloud services evaluation domain, we plan to unfold our future work along two directions. Firstly, we will keep improving DoKnowMe by further specifying and supplementing more strategies for individual evaluation steps and activities. In contrast to our defined domain-specific knowledge artefacts, the common strategies can be viewed as domain-independent knowledge of conducting performance evaluation. Secondly, we are about to apply DoKnowMe to the performance evaluation of other software systems. This work will in turn help validate DoKnowMe in different system domains.


\bibliographystyle{abbrv}
\bibliography{PER_Ref}  

\begin{thebibliography}{10}

\bibitem{Bowen_Kreuter_2009}
D.~J. Bowen, M.~Kreuter, B.~Spring, L.~Cofta-Woerpel, L.~Linnan, D.~Weiner,
  S.~Bakken, C.~P. Kaplan, L.~Squiers, C.~Fabrizio, and M.~Fernandez.
\newblock How we design feasibility studies.
\newblock {\em Am. J. Preventive Med.}, 36(5):452--457, May 2009.

\bibitem{Checkland_Scholes_1999}
P.~Checkland and J.~Scholes.
\newblock {\em Soft Systems Methodology in Action}.
\newblock John Wiley \& Sons Ltd., New York, NY, September 1999.

\bibitem{Denning_1981}
P.~J. Denning.
\newblock Performance evaluation: Experimental computer science at its best.
\newblock {\em ACM SIGMETRICS Perform. Eval. Rev.}, 10(3):106--109, Fall 1981.

\bibitem{Feitelson_2006}
D.~G. Feitelson.
\newblock Experimental computer science: The need for a cultural change.
\newblock \url{http://www.cs.huji.ac.il/~feit/papers/exp05.pdf}, 3 December
  2006.

\bibitem{Feitelson_2015}
D.~G. Feitelson.
\newblock From repeatability to reproducibility and corroboration.
\newblock {\em ACM SIGOPS Operating Syst. Rev.}, 49(1):3--11, January 2015.

\bibitem{Fortier_Michel_2003}
P.~J. Fortier and H.~E. Michel.
\newblock {\em Computer Systems Performance Evaluation and Prediction}.
\newblock Digital Press, Burlington, MA, July 2003.

\bibitem{Grossklags_2007}
J.~Grossklags.
\newblock Experimental economics and experimental computer science: A survey.
\newblock In {\em Proc. ExpCS 2007}, pages 1--11, San Diego, CA, USA, 13-14
  June 2007. ACM Press.

\bibitem{Heiser_2014}
G.~Heiser.
\newblock Systems benchmarking crimes.
\newblock \url{http://www.cse.unsw.edu.au/~gernot/benchmarking-crimes.html}, 13
  March 2015.

\bibitem{Islam_Lee_2012}
S.~Islam, K.~Lee, A.~Fekete, and A.~Liu.
\newblock How a consumer can measure elasticity for {C}loud platforms.
\newblock In {\em Proc. ICPE 2012}, pages 85--96, Boston, USA, 22-25 April
  2012. ACM Press.

\bibitem{Jain_1991}
R.~K. Jain.
\newblock {\em The Art of Computer Systems Performance Analysis: Techniques for
  Experimental Design, Measurement, Simulation, and Modeling}.
\newblock Wiley Computer Publishing, John Wiley \& Sons, Inc., New York, NY,
  April 1991.

\bibitem{Le_Boudec_2011}
{Jean-Yves Le Boudec}.
\newblock {\em Performance Evaluation of Computer and Communication Systems}.
\newblock EFPL Press, Lausanne, Switzerland, February 2011.

\bibitem{Koziolek_2010}
H.~Koziolek.
\newblock Performance evaluation of component-based software system: {A}
  survey.
\newblock {\em Perform. Eval.}, 67(8):634–658, August 2010.

\bibitem{Li_OBrien_2013_boosting}
Z.~Li, L.~O'Brien, R.~Cai, and H.~Zhang.
\newblock Boosting metrics for {Cloud} services evaluation -- the last mile of
  using benchmark suites.
\newblock In {\em Proc. AINA 2013}, pages 381--388, Barcelona, Spain, 25-28
  March 2013. IEEE Computer Society.

\bibitem{Li_OBrien_2012_factor}
Z.~Li, L.~O'Brien, H.~Zhang, and R.~Cai.
\newblock A factor framework for experimental design for performance evaluation
  of commercial {Cloud} services.
\newblock In {\em Proc. CloudCom 2012}, pages 169--176, Taipei, Taiwan, 3-6
  December 2012. IEEE Computer Society.

\bibitem{Li_OBrien_2012_metrics}
Z.~Li, L.~O'Brien, H.~Zhang, and R.~Cai.
\newblock On a catalogue of metrics for evaluating commercial {C}loud services.
\newblock In {\em Proc. GRID 2012}, pages 164--173, Beijing, China, 20-23
  September 2012. IEEE Computer Society.

\bibitem{Li_OBrien_2014}
Z.~Li, L.~O'Brien, H.~Zhang, and R.~Cai.
\newblock On the conceptualization of performance evaluation of {IaaS}
  services.
\newblock {\em IEEE Trans. Serv. Comput.}, 7(4):628--641, October-December
  2014.

\bibitem{Li_Zhang_2013_slr}
Z.~Li, H.~Zhang, L.~O'Brien, R.~Cai, and S.~Flint.
\newblock On evaluating commercial {Cloud} services: {A} systematic review.
\newblock {\em J. Syst. Softw.}, 86(9):2371--2393, September 2013.

\bibitem{Montgomery_2009}
D.~C. Montgomery.
\newblock {\em Design and Analysis of Experiments}.
\newblock John Wiley \& Sons, Inc., Hoboken, NJ, 7th edition edition, January
  2009.

\bibitem{Obaidat_Boudriga_2010}
M.~S. Obaidat and N.~A. Boudriga.
\newblock {\em Fundamentals of Performance Evaluation of Computer and
  Telecommunication Systems}.
\newblock John Wiley \& Sons, Inc., Hoboken, New Jersey, January 2010.

\bibitem{Osman_Knottenbelt_2012}
R.~Osman and W.~J. Knottenbelt.
\newblock Database system performance evaluation models: {A} survey.
\newblock {\em Perform. Eval.}, 69(10):471--493, October 2012.

\bibitem{Price_Baecker_1993}
B.~A. Price, R.~M. Baecker, and I.~S. Small.
\newblock A principled taxonomy of software visualization.
\newblock {\em J. Visual Lang. Comput.}, 4(3):211--266, September 1993.

\bibitem{Runeson_2009}
P.~Runeson and M.~H{\"{o}}st.
\newblock Guidelines for conducting and reporting case study research in
  software engineering.
\newblock {\em Empir. Softw. Eng.}, 14(2):131--164, April 2009.

\bibitem{Kistowski_Herbst_2014}
J.~v.~Kistowski, N.~Herbst, and S.~Kounev.
\newblock Modeling variations in load intensity over time.
\newblock In {\em Proc. LT 2014}, pages 1--4, Dublin, Ireland, 22 March 2014.
  ACM Press.

\end{thebibliography}
\end{document}